\documentclass[aps,prl,preprintnumbers,epsf,nofootinbib]{revtex4}
\usepackage[dvips]{graphicx}
\newcommand{\bea}{\begin{eqnarray}}
\newcommand{\eea}{\end{eqnarray}}

\begin{document}
\preprint{arXiv: 0901.3731v4[hep-th]}
\title{A note on nonperturbative renormalization of effective field theory}
\author{Ji-Feng Yang}
\address{Department of Physics, East China Normal University,
Shanghai, 200062, China}
\begin{abstract}
Within the realm of contact potentials, the key structures intrinsic
of nonperturbative renormalization of $T$-matrices are unraveled using
rigorous solutions and an inverse form of algebraic Lippmann-schwinger
equation. The intrinsic mismatches between effective field theory power
counting and nonperturbative divergence structures are shown for the
first time to preclude the conventional counterterm algorithm from
working in the renormalization of EFT for $NN$ scattering in
nonperturbative regimes.
\end{abstract}
\maketitle
\section{Introduction}
The effective field theory (EFT) approach to nucleon systems has
been producing many encouraging results\cite{BvKERev}, pointing
towards a promising field-theoretical framework for nuclear forces.
In this course, evidences have been accumulated that the
conventional renormalization programs cease to apply in a
straightforward manner for such nonperturbative problems, along with
debates concerning this issue\cite{NTvK,EpelMeis}. This is not
totally unexpected as the issue is nonperturbative which may pervert
the wisdoms established within perturbative frameworks. For example,
as noted in Ref.\cite{kaplan}, perturbative analysis of
counterterms\cite{KSW} can be misleading, therefore "new theoretical
ideas" for nonperturbative treatment of EFT are needed. The
nonperturbative aspects of this issue are also emphasized in
\cite{LvK}.

Actually, the difficulties encountered so far even brought about
some doubts concerning the validity of EFT approach to the nuclear
systems. In our view, it is natural to think of field-theoretical
approach to nuclear systems as an important advancement, while the
difficulties imply that such treatment has not been fully
accomplished yet. Therefore, it is important to unravel hidden
structures and notions underlying the nonperturbative
renormalization of EFT for nucleon-nucleon ($NN$) interactions. For
this purpose, we will work with contact potentials or pionless EFT
to obtain rigorous solutions that could make things transparent.
Here, we remind that our main purpose here is not to reproduce the
well known results about pionless EFT in literature, say,
\cite{KSW,vK99}, but to explore the nonperturbative properties or
structures that should be generally useful for nonperturbative
renormalization of both pionless and pionful EFT's, and even other
effective theories.
\section{Parametrization and rigorous solutions}
The setup is as follows: The potentials for $NN$ scattering are
first systematically constructed using chiral perturbation theory
($\chi$PT) up to some chiral order $\Delta$ and then resummed
through Lippmann-Schwinger equations (LSE's) to obtain the
$T$-matrices\cite{WeinEFT}. In case of contact potentials, the LSE's
could be turned into algebraic ones using following trick or
ansatz\cite{Phillips,C71} (we consider an uncoupled partial wave
channel $L$ for simplicity): \bea
&&V_L(q,q^\prime)=q^L{q^\prime}^L\sum_{i,j=0,1,2,
\cdots}\lambda_{ij}q^{2i}{q^\prime}^{2j}=q^L{q^\prime}^LU^T(q^2)\lambda
U({q^\prime}^2),\\ &&T_L(q,q^\prime)=q^L{q^\prime}^L\sum_{i,j=0,1,2,
\cdots}\tau_{ij}q^{2i}{q^\prime}^{2j} =q^L{q^\prime}^LU^T(q^2)\tau
U({q^\prime}^2),\eea with $q,q^\prime$ being external momenta and
$U^T(q^2)\equiv(1,q^2,q^4,\cdots)$. Here $\lambda$ is
energy-independent while $\tau$ is energy-dependent\footnote{It is
known that energy dependence in the potentials could be removed
through unitary transformations, see e.g., Ref.\cite{EGM98}.}. As
$V_L$ is truncated at a finite order $\Delta$ according to EFT power
counting, we have the following constraints:\bea\label{constraint}
\lambda_{ij}=0,\ \forall\ i,j:\ i+j>\Delta/2-L.\eea This constraint
will prove to be crucial. The algebraic LSE for channel $L$ now
reads, \bea \label{LSE-algebra} \tau(E)&&=\lambda+\lambda
\circ{\mathcal{I}}(E)\circ\tau(E),\eea with \bea
{\mathcal{I}}(E)\equiv({\mathcal{I}}_{ij}(E)), {\mathcal{I}}_{ij}(E)
\equiv\int
\frac{d^3k}{(2\pi)^3}\frac{k^{2(i+j)}}{E-k^2/M+i\epsilon},\
i,j=0,1,2\cdots.\eea So, the renormalization of $T$'s boils down to
the renormalization of $\tau$'s as $U(q^2)$ or $U^T(q^2)$ is not
subject to renormalization at all. Our analysis here are also
illuminating for the more realistic cases with pion exchanges, as
the LSE there is still dominated by power like divergences:
$V(q,q^\prime,\cdots)\sim \sum q^\alpha{q^\prime}^\beta$ when
$q,q^\prime \rightarrow\infty$. We note in passing that the above
theoretical setup may also be applied to other problems dominated by
singular short range interactions.

Now we parametrize the divergent integrals $[\mathcal{I}_{ij}(E)]$
in the following general manner:
$\mathcal{I}_{ij}(E)\equiv\sum_{m=1}^{i+j}
J_{2m+1}p^{2(n-m)}-{\mathcal{I}}_0p^{2(i+j)}$, where $p=\sqrt{ME}$
and ${\mathcal{I}}_0\equiv J_0+i\frac{Mp}{4\pi}$ and the arbitrary
parameters $J_0$ and $J_{2m+1} (m=1,2,\cdots)$ parametrize any
sensible regularization/renormalization scheme. Generically, $J_0$
and $J_{2m+1}$ should be independent of energy. Then
${\mathcal{I}}(E)$ takes the following form in $^1S_0$ channel
($L=0$), \bea\label{I-para-1s0} {\mathcal{I}}(E)&&\equiv
-\mathcal{I}_0 U(p^2)U^T(p^2) +J_3\Delta U_1(p^2)+J_5\Delta
U_2(p^2)+\cdots,\eea with \bea\Delta
U_1(p^2)\equiv\frac{1}{p^2}\int^{p^2}_0 dt
\frac{d[U(t)U^T(t)]}{dt},\ \Delta U_{n+1}(p^2)\equiv
\frac{1}{p^2}\int^{p^2}_0 dt \frac{d[\Delta U_n(t)]}{dt}, \
n\geq1.\eea While for $L\geq1$, we have, \bea\label{I-para-L}
{\mathcal{I}}(E)=\left(-\mathcal{I}_0p^{2L}+J_3p^{2L-2}+
\cdots+J_{2L+1} \right)U(p^2)U^T(p^2) +J_{2L+3}\Delta U_1(p^2)
+J_{2L+5}\Delta U_2(p^2)+\cdots.\eea Obviously, any sensible
prescription could be readily reproduced by assigning appropriate
values to $J_{\cdots}$.

Some remarks are in order: First, all the divergent integrals
involved assemble into the matrix ${\mathcal{I}}(E)$ of finite rank, or
finite many divergences are to be treated. This 'finiteness' should
be able to substantiate the nonperturbative renormalization of $T$.
Second, the parameters $[J_{\cdots}]$ in ${\mathcal{I}}(E)$ are
nonperturbative and irreducible in the sense that they will appear
as basic parameters in the compact form of $T$ . Third, $J_0$ is
very special as it always appears together with $i\frac{Mp}{4\pi}$
in each entry of ${\mathcal{I}}(E)$ while $[J_{2m+1}, m>0]$ do not.

The algebraic LSE could now be readily solved (the energy-dependence
in $\tau$ and $\mathcal{I}$ will be omitted below):
\bea\label{algtau}\tau=(1-\lambda\circ{\mathcal{I}}\circ)^{-1}
\lambda=\lambda(1-\circ{\mathcal{I}}\circ\lambda)^{-1}.\eea Then the
on-shell $T$ for channel $L$ reads\cite{C71} (from now on we use
$[C_{\cdots}]$ to denote $[\lambda_{\cdots}]$):\bea \label{InvT-ll}
\frac{1}{{\bf T}_L}\equiv\frac{1}{T_L(q,q^\prime)}|_{q=q^\prime=p}
=\mathcal{I}_0 +\frac{ N_L([C_{\cdots}],[J_{2m+1}],p^2)}
{D_L([C_{\cdots}],[J_{2m+1}],p^2)p^{2L}},\eea where $N_L$ and $D_L$
are polynomials in terms of real parameters: the contact couplings
$[C_{\cdots}]$, $[J_{2m+1},m>0]$ and $p^2$. While for coupled
channels ($^3L_{J}{-} ^3L^\prime_J, L=J-1,L^\prime=J+1$), one could
find the following\cite{3s1-3d1}:\bea \label{TINV}{\bf
T}^{-1}_J={\mathcal{I}_0}\left(\begin{array}{cc}1&0\\0&1\\\end{array}
\right) + \left(\begin{array}{cc}
\frac{\mathcal{N}_{L,L}}{\mathcal{D}_{L,L}p^{2L}} ,&
\frac{-\mathcal{N}_{L,L^\prime}}{\mathcal{D}_{L,L^\prime}p^{2J}}\\
\frac{-\mathcal{N}_{L,L^\prime}}{\mathcal{D}_{L,L^\prime}p^{2J}}, &
\frac{\mathcal{N}_{L^\prime,L^\prime}}
{\mathcal{D}_{L^\prime,L^\prime} p^{2L^\prime}}\\\end{array}
\right).\eea Again $[\mathcal{N}_{\cdots},\mathcal{D}_{\cdots}]$ are
real polynomials in terms of $[C_{\dots}]$, $[J_{2m+1},m>0]$ and
$p^2$. Note that such $T$-matrices are automatically unitary.
\section{Renormalization of effective field theories in nonperturbative regimes}
In the following, it suffices to mainly work with the uncoupled
channels for unraveling the novel features of renormalization that
elude the conventional perturbative scenario and wisdoms.
\subsection{On-shell cases}
First, let us consider the on-shell case. The on-shell $T$-matrices
given in Eqs.(\ref{InvT-ll}) and (\ref{TINV}) exhibit the following
important features worth emphasis\cite{C71,4637,3s1-3d1}: (1) First,
the same complex parameter $\mathcal{I}_0$ appears in all channels
in the same isolated position in $1/T$ or ${\bf T}^{-1}$, i.e.,
$\mathcal{I}_0$ is 'decoupled' from $[C_{\cdots}]$ and $[J_{2m+1},m>0]$
in every channel\footnote{The rigorous proof of this point for $^1S_0$
channel has been given in Ref.\cite{C71}, which could be generalized to
higher channels.}. (2) Second, as is already noted above, only finite
many irreducible divergences $[J_{\cdots}]$ enter the game. That is,
$\text{Rank}({\mathcal{I}})<\infty$.

Since the $p$-dependence of the on-shell $T$-matrices is physical,
the prescription variations (i.e., variations in $[J_{\cdots}]$)
must be compensated by that of the couplings. This is nothing else
but the principle of RG invariance, then appropriate combinations of
$[N_{\cdots}]$ and $[D_{\cdots}]$ must be RG invariants. Moreover,
the isolation of $\mathcal{I}_0$ in all $T^{-1}$'s makes it alone a
RG invariant parameter to be physically
determined\cite{C71,3s1-3d1,4637}. Therefore, {\em in
nonperturbative regime}, $J_0$ becomes a universal physical scale in
the low energy $NN$ scattering. This is not a bizarre event: In
Wilsonian approach, the nontrivial fixed-point solution\cite{Birse}
just equals to the negative inverse of $J_0$ computed in cutoff
scheme: $(\hat{V}_0(p))^{-1}=\frac{M}{2\pi^2} \left(-\Lambda+
\frac{p}{2}\ln\frac{\Lambda+p}{\Lambda-p}\right)
=-J_0(p,\Lambda)=-\text{Re} (\mathcal{I}_{0;\Lambda})$. There is
only one exception at leading order in $^1S_0$ where $J_0$ mixes
with $1/C_0$\cite{KSW}.

There are also some cases at lower orders where some divergences in
$[J_{2m+1}, m\neq0]$ might be absorbed into the couplings. For
example, at $\Delta=2$, the inverse on-shell $T$ for
$^1S_0$\cite{Phillips} reads
\bea\frac{1}{T}=\mathcal{I}_0+\frac{N_0}{D_{0;0}+D_{0;1}p^2}, \eea
with \bea N_0=(1-C_2J_3)^2,\ D_{0;0}=C_0+C_2^2 J_5,\
D_{0;1}=C_2(2-C_2J_3).\eea It could be make finite by requiring
$\frac{D_{0;0}}{N_0}$ and $\frac{D_{0;1}}{N_0}$ to be finite
constants: $\frac{1}{T}=\mathcal{I}_0+\frac{1}{c_{0}+2c_{2}p^2}$.
The solutions are quite sophisticated\cite{C71},\bea
C_2^{(\pm)}=J_3^{-1}\left(1\pm (1+2c_{2}J_3)^{-1/2}\right),\quad
C_0^{(\pm)}=\frac{c_{0}}{1+2c_{2}J_3}- \frac{J_5}{J^2_3}\left(1 \pm
(1+2c_{2}J_3)^{-1/2} \right)^2.\eea However, there is no way to
subtract the divergence in $J_0$ with such sophisticated
counterterms, thus the counterterm algorithm failed here.

Things get worse at higher orders. For example, at $\Delta=4$ for
$^1S_0$, we have\cite{C71}, \bea\frac{1}{T}=\mathcal{I}_0
+\frac{N_0+N_1p^2+N_2p^4}{D_0+D_1p^2+D_2p^4+D_3p^6}\eea with $
N_2=C_4^2J_3^2,\ D_3=-C_4^2J_3$, where it is simply impossible to
renormalize $N_2$ and $D_3$ with counter terms from couplings at the
same time as $N_2/D_3=-J_3$ contains a divergence! This status is
generically true at higher orders, regardless of channels. Then, in
order to obtain finite results, we have to go beyond the
counterterms from couplings. All these points hint us about
something unprecedented. To unravel them, we turn to the off-shell
case.
\subsection{Off-shell cases}
As already pointed out above, it suffices to consider the
renormalization of $\tau$. To expose the most crucial
nonperturbative structures, let us turn Eq.(\ref{algtau}) into the
following inverse form, \bea \label{inv-algebraLSE} \tau^{-1}=
\lambda^{-1}-{\mathcal{I}},\eea in terms of which the unitarity $
\tau^{-1}-(\tau^\dagger)^{-1}=\frac{iMp}{2\pi}U(p)U^T(p)$ is
obviously not affected by renormalization at all. Since the $p$
dependence of $T$ is physical, so is it for $\tau$. Thus $\tau$
must be finite and prescription-independent. Then
Eq.(\ref{inv-algebraLSE}) tells us that the renormalization of $\tau$
ultimately reduces to the removal of divergences in ${\mathcal{I}}$
in such a manner that the combination $\lambda^{-1}-{\mathcal{I}}$ is
RG invariant or physical.

At first sight, this seems trivial as one could let $\lambda^{-1}$
absorb all the divergences, i.e., counterterms from $\lambda^{-1}$ is
at work. Unfortunately, this is not true due to the following two
intrinsic mismatches between $\lambda^{-1}$ and ${\mathcal{I}}$: (i)
$\lambda^{-1}$ is constrained as follows due to the truncation
constraint (\ref{constraint}):
\bea\label{invconstraint}(\lambda^{-1})_{ij}=0,\ \forall\ i,j:\
i+j\le \Delta/2-L,\eea while ${\mathcal{I}}$ is free from such
constraints; (ii) ${\mathcal{I}}$ is energy-dependent while
$\lambda^{-1}$ is not.

Let us elaborate. According to EFT power counting, the counterterms
must also be constrained by Eq.(\ref{invconstraint}). Then there is
no counterterm for the entries ${\mathcal{I}}_{ij}$ where
$(\lambda^{-1})_{ij}=0$ to subtract the divergences there, that is,
the counterterm algorithm could not perform sufficient subtractions.
For example, for $^1S_0$ at $\Delta=2$, we have
\bea\lambda=\left(\begin{array}{cc} C_0 & C_2\\ C_2&0 \\
\end{array}\right)\Rightarrow \lambda^{-1}=\left( \begin{array}{cc}
0& C^{-1}_2\\C^{-1}_2 & -C_0C_2^{-2} \\\end{array}\right).\eea In
the meantime, \bea {\mathcal{I}}=\left( \begin{array}{cc}
-{\mathcal{I}}_0 & J_3-{\mathcal{I}}_0p^2 \\
J_3-{\mathcal{I}}_0p^2& J_5+J_3p^2-{\mathcal{I}}_0p^4\\
\end{array}\right).\eea It is obvious that the divergence in
$\mathcal{I}_{0,0}$ (i.e., in $J_0$) could by no way be subtracted
by counterterms from $(\lambda^{-1})_{0,0}$, which is zero as
required by consistent EFT power counting. There is no sensible way
within nonperturbative regimes to remove such inherent mismatches
between EFT power counting and the divergence 'configuration', hence
the counterterms from EFT couplings could not work here. This can be
seen as following: Suppose we introduce higher order terms in
potential so that the mismatch is gone, for the example considered
above, it means $\lambda_{1,1}\neq0$ or the term $\sim
q^2{q^\prime}^2$ is included and hence $(\lambda^{-1})_{0,0}\neq0$,
then the general principle of EFT power counting is broken,
according to which terms $\sim q^4$ and $\sim{q^\prime}^4$ should
also be included, which means $\lambda_{2,0}=\lambda_{0,2}\neq0$. It
will not help by including $\lambda_{2,0}$ and $\lambda_{0,2}$ as
then $\lambda$ and hence ${\mathcal{I}}$ will be enlarged, and the
mismatch will persist between the enlarged $\lambda^{-1}$ and
${\mathcal{I}}$, unless one 'removes' the truncation itself, which
is actually impossible in any EFT approach.

It will also not help even one ignores the EFT power counting in
constructing counterterms, since Eq.(\ref{inv-algebraLSE}) means the
counterterms would necessarily develop energy-dependence as the
divergences in ${\mathcal{I}}$ are energy dependent, which is
theoretically unfavorable. This is due to the fact that the
"nonperturbative" counterterms introduced through $\lambda^{-1}$
would lead to non-polynomial energy dependence in the local
couplings, that is, the operators that are "nonlocal" in time, not
the "local" ones that are allowed within contact potential approach.
In other words, even one works with energy dependent version of
potentials (or operators "local" in time), there still may appear
mismatches between the inverse $\lambda^{-1}$, which is now
non-polynomial in terms of energy, and the integrals
${\mathcal{I}}$, which is polynomial in terms of energy. Therefore,
such choices could not remove the mismatches unraveled, actually, it
lead to new serious problems. As the mismatches only originated from
the nonperturbative structures of the divergences involved, we
suspect that they may also persist in the more realistic cases with
nonlocal potentials, especially for the cases with pion exchanges
that interests most practitioners in the EFT approach to nuclear
forces. Of course, definite conclusions are not available before
rigorous solutions of such cases are available.

At this stage, we note that there is an exception at leading order
($\Delta=0$) where $\lambda=C_0, {\mathcal{I}}=-{\mathcal{I}}_0$ and the
mismatches between $\lambda^{-1}$ and ${\mathcal{I}}$ are gone. Since
the divergence status is not altered after one-pion-exchange is included,
this could explain why counterterm algorithm works in such
cases\cite{NTvK,YPhillips,EPVRAM}.

The intrinsic relations between EFT power counting and nonperturbative
structures of divergences naturally leads us to conclude that, beyond
leading order, it is generally impossible to implement the counterterms
from couplings in nonperturbative regimes. However, this is not
equivalent to the failure of renormalization itself. In this connection,
we recall that the ultimate goal of renormalization is to obtain finite
amplitudes generated with EFT propagators and vertices, not how the
divergences are removed, and that the most crucial step is to fix the
undetermined constants generated in any renormalization procedure by
imposing appropriate boundary conditions. Due to the difficulties
described above, one is naturally led to the subtraction directly
performed on the integrals in ${\mathcal{I}}$, or through other means
that could yield equivalent effects. In whatever means, the final outcome
is that, due to the energy or $p$ dependence of $\mathcal{I}$ (C.f.
Eq.(\ref{I-para-1s0}) or (\ref{I-para-L})), at least the constant $J_0$
(which is energy- or $p$-independent) could no longer mix with the
couplings in $\lambda^{-1}$ and therefore must be physically determined
through imposition of appropriate boundary conditions. Thus the two
mismatches between $\lambda^{-1}$ and $\mathcal{I}$ lead to the RG
invariance of $J_0$. In fact, as long as a $J_{2m+1} (m>0)$ appears as
coefficient of a $p$-dependent matrix $U(p^2)U^T(p^2)$ or $\Delta
U_n(p^2)(n\geq1)$, it must also be physically determined. Thus we
reproduced the same conclusion as obtained in the on-shell case.
\section{Discussions and summary}
Thus through the above analysis within the realm of contact potentials
or EFT($\not\!\pi$), we showed that the conventional counterterm algorithm
and the associated wisdoms could not work beyond the leading order due
to the intrinsic relations between EFT power counting and nonperturbative
divergences. Then one must resort to other approaches beyond the
counterterm algorithm. Here, we note that the counterterm algorithm
refers to the construction of counterterms from EFT vertices or
potentials, not the subtraction in general sense. Thus, in the
treatments of $NN$ scattering at higher orders where rigorous and
explicit parametrization of the divergences is impossible and
counterterms could not work, keeping the cutoff finite (which is one
kind of subtraction already) and properly fine-tuning it together with
other contact couplings is a choice that is pragmatic and
reasonable\cite{EGM,EMach}. Or, one may choose some 'perturbative'
treatment of the potentials beyond leading order as long as
the convergence is assured\cite{BKV}. For further progresses, the
nonperturbative properties and mismatches revealed here should be
illuminating and hence carefully taken into account.

Before closing our presentation, let us expose another interesting
point associated with inverse formalism that is intrinsically
nonperturbative: The EFT power counting expressed in terms of
$\lambda^{-1}$ seems somewhat 'unusual' due to the constraint
(\ref{invconstraint}): There are some entries that jump to zero and
deviate from the seemingly well ordered sequence of the nonzero entries.
This is again due to the truncation of potential which is natural from
EFT side. Further exploration of this point will be pursued in the
future.

In summary, we provided a somewhat transparent analysis of the
renormalization of EFT in nonperturbative regimes within the context
of contact potentials or pionless EFT without introducing any
deformation of the standard field-theoretical framework. In a
formulation that makes the main structural issues lucid and obvious,
it was shown for the first time that the intrinsic mismatches
between EFT power counting and nonperturbative divergences preclude
counterterm algorithms from being at work. Possible ways out and the
reasonable aspects of some approaches were also briefly addressed.
The notions revealed here could well be applied to wider range of
physical systems that are dominated by singular short-distance
interactions.
\section*{Acknowledgement}
The project is supported in part by the National Natural Science
Foundation under Grant No. 10205004 and by the Ministry of Education of
China.

\end{document}